\documentclass[aip,apl,reprint,  showpacs,showkeys]{revtex4-1}

\usepackage{graphicx}
\usepackage{amsmath,amssymb}
\usepackage{bm}
\usepackage{color}

\begin{document}

\title{Laser-induced quantum pumping in graphene}

\author{Pablo San-Jose}
\affiliation{Instituto de Estructura de la Materia (IEM-CSIC), Serrano 123, 28006 Madrid, Spain}

\author{Elsa Prada}
\affiliation{Instituto de Ciencia de Materiales de Madrid, CSIC, Cantoblanco, 28049 Madrid, Spain}

\author{Henning Schomerus}
\affiliation{Department of Physics, Lancaster University, Lancaster, LA1 4YB, United Kingdom}

\author{Sigmund Kohler}
\affiliation{Instituto de Ciencia de Materiales de Madrid, CSIC, Cantoblanco, 28049 Madrid, Spain}

\date{\today}

\begin{abstract}
We investigate non-adiabatic electron pumping in graphene generated by
laser irradiation with linear polarization parallel or perpendicular to
the transport direction. Transport is dominated by the spatially
asymmetric excitation of electrons from evanescent into propagating
modes. For a laser with parallel polarization, the pumping response
exhibits a subharmonic resonant enhancement which directly probes the
Fermi energy; no such enhancement occurs for perpendicular polarization.
The resonance mechanism relies on the chirality of charge carriers in
graphene.
\end{abstract}
\maketitle

The periodic modulation of a mesoscopic conductor attached to electronic
reservoirs can be used to induce a transfer of electrons across the
system, which produces a dc current in absence of a bias voltage. Such
\emph{quantum pumps} can be driven adiabatically slowly, provided that
their work cycle is oriented,\cite{Brouwer1998a} which requires
variation of at least two parameters. For faster, non-adiabatic driving,
single-parameter pumping is possible,\cite{Kohler2005a,Torres:APL11,Zhou:12}
which has been observed in a number of experiments.\cite{Oosterkamp1998a,
Blumenthal2007a, DiCarlo2003a, Kaestner2008a, Fujiwara2008a,
Kaestner2009a, Khrapai2006a}

\begin{figure}[b]
\includegraphics[width=\columnwidth]{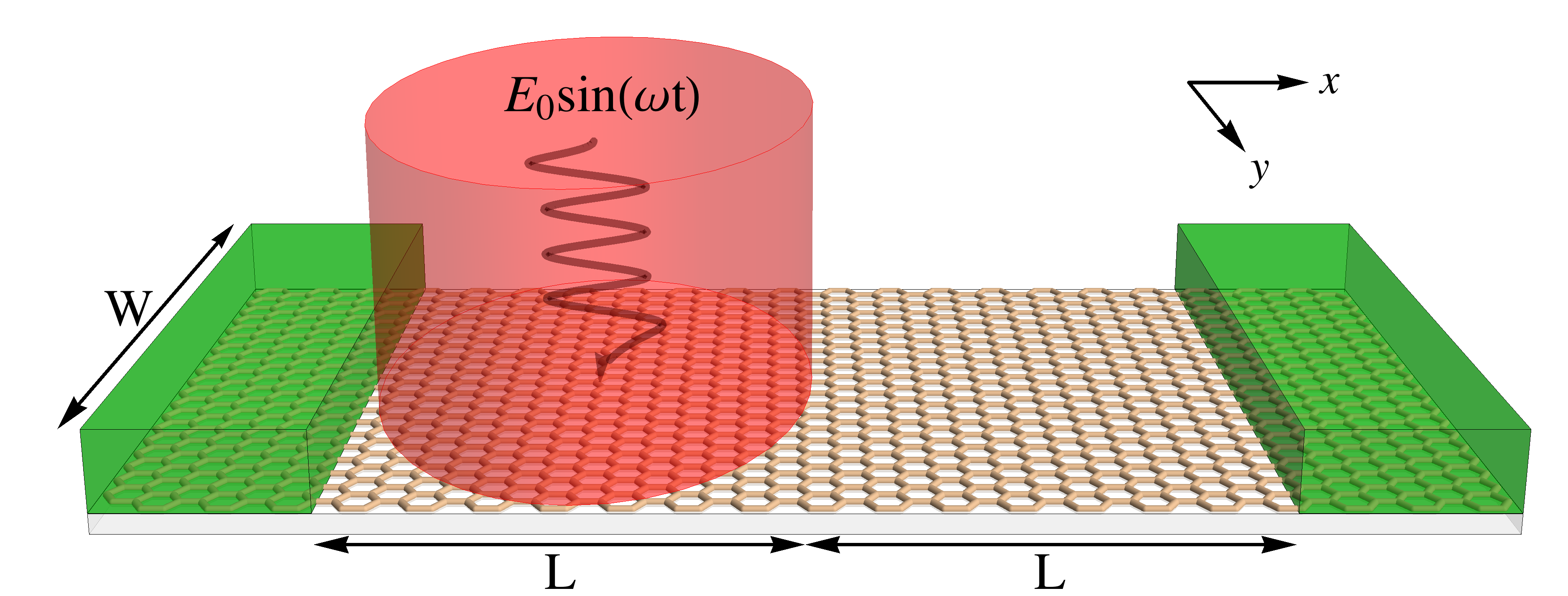}
\caption{Sketch of a graphene sheet of length $2L$ and width $W$
connected to two highly doped graphene leads. The left half of the
sheet is driven periodically in time by laser irradiation of frequency
$\omega$ and amplitude $E_0$.
}
\label{fig:model}
\end{figure}

With the advent of graphene, it has been suggested to exploit and probe
the properties of this material both by adiabatic \cite{Prada2009a,
Wakker2010a, AlosPalop2011a, Liu:N11,Low2012a} and non-adiabatic pumping,
\cite{SanJose2011a, Kaur2012a} with the driving typically provided by
applying ac electrostatic voltages. In this work, by contrast, we
consider driving graphene by linearly polarized laser irradiation,
\cite{Calvo:APL11,Gu:PRL11} as is depicted in Fig.~\ref{fig:model}. We show
that this results in a sizable pumped current, which moreover is
resonantly enhanced when the polarization of the electric field is
parallel with the transport direction. The resonance occurs at a Fermi
energy equaling half the photon energy, and thus may be useful to
characterize the properties of a graphene sample. The polarization
dependence arises from a selection rule which directly probes the
chirality of the charge carriers in graphene.

We consider a graphene monolayer sheet of length $2L$ and width $W$
connected to two highly doped leads, as shown in Fig.~\ref{fig:model}. In
the absence of the laser irradiation, the potential energy in the sheet
is $U_B(x)=0$, while in the leads $U_B(x)\ll 0$; the potential energy is
independent of the transverse coordinate $y$. The low-energy excitations
are then described by the quasi-one-dimensional Dirac Hamiltonian $H_0 =
\hbar v_F \bm k\cdot\bm\sigma + U_B(x)$, where $v_F \approx
10^6\,\mathrm{m/s}$ denotes the Fermi velocity of graphene. The
components of $\bm\sigma$ are the Pauli matrices $\sigma_x$ and
$\sigma_y$, which act in a pseudo-spin space that physically describes
the amplitudes on the two inequivalent lattice sites of graphene's
honeycomb lattice. Because of the term $\bm{k}\cdot\bm{\sigma}$ in $H_0$,
the electronic states of graphene possess chirality, which is responsible
for many of its properties,\cite{CastroNeto2009a} including those that
result in efficient pumping by ac gating.\cite{Prada2009a, SanJose2011a}

The pumping is induced by laser irradiation focused on a region of size
$L$ next to the left electrode, with linear polarization along the unit
vector $\bm n$ in the plane of the graphene sheet. The associated
oscillating electric field $\bm{E}(x,t)=E_0\sin(\omega t)a(x) \bm n$ is
modeled by a piecewise constant laser spot profile $a(x)$ (the magnetic
field in the graphene plane has a negligible physical effect). To avoid
complications arising from a position-dependent scalar potential, we work
with the Weyl gauge in which the electric field is derived from a
time-dependent vector potential, $\bm E (x,t)= -\partial{\bm
A(x,t)}/\partial t$. Using the minimal coupling $\bm k\to \bm k+e\bm A$
for the electron charge $-e$, we obtain the time-dependent Dirac
Hamiltonian
\begin{equation}\label{H}
H(t) = H_0 + a(x) \cos(\omega t) U_{\bm{n}},
\end{equation}
where the constant coupling operator is given by
\begin{eqnarray}
\label{Vlaser}
U_{\bm{n}}&=&ev_F\frac{E_0}{\omega}\bm{n}\cdot \bm\sigma.
\end{eqnarray}
It follows that the polarization direction $\bm{n}$ couples to the
pseudo-spin $\bm{\sigma}$ and thus affects the quasiparticle chirality.

The pumped current can be obtained by Floquet scattering theory,
\cite{Moskalets:PRB02a} which is valid as along as interactions can be
neglected. At low temperatures, the dc current is then given by the
generalized Landauer formula \cite{Wagner1999a,Kohler2005a}
\begin{equation}
\label{current}
I = \frac{4e}{h}\int^{E_F}_{-\infty} d\epsilon \,\Delta T(\epsilon),
\end{equation}
where the factor 4 accounts for spin and valley degeneracy. The central
quantity is the net transmission imbalance $\Delta T(\epsilon) = \sum_n
[T_{LR}^{(n)}(\epsilon) -T_{RL}^{(n)}(\epsilon)]$, which involves the
transmission coefficients $T_{\ell\ell'}^{(n)}(\epsilon)$ from lead
$\ell=L,R$ to lead $\ell'\neq\ell$, where $n$ labels sidebands reached by
absorption or emission of $n$ photons ($n<0$ corresponds to emission).
The definition of the transmission coefficients includes the summation
over the transverse wave number $k_y$, which here is conserved, and can be
considered continuous for wide and short pumps ($W\gg L$). The pumped
current is non-zero since the specified setup is asymmetric. In
particular, as in the case of driving by an ac gate voltage,
\cite{Prada2009a, SanJose2011a} carriers entering the driving region
from the left lead in evanescent modes can be promoted to propagating
modes, which allows them to reach the right lead, while electrons
arriving in evanescent modes from the right lead practically never reach
the driven region; the result is net transport from left to right.

Before presenting quantitative results we establish the preferred pumping
regime and the typical magnitude of the pumped current. There are two
relevant parameters. The first is the effective driving strength
\begin{equation}
p=\left(\frac{|U_{\bm{n}}|}{2\hbar\omega}\right)^2
=\left(\frac{ev_FE_0}{2\hbar\omega^2}\right)^2.
\label{eq:p2}
\end{equation}
Adjacent sidebands possess relative energy shifts $\hbar\omega$, and are
coupled by a ``hopping'' energy $|U_{\bm{n}}|/2$. The ratio of these two
quantities, $\sqrt{p}$, equals the number of sidebands that get populated
in the scattering process,\cite{SanJose2011a} i.e., the average number
of absorbed or emitted photons. For $p\ll1$, only one-photon processes
play a practical role. Within the single-sideband approximation $\Delta
T$ is proportional to $p$, so that for small $p$ the net transmission
$\Delta T/p$ is independent of driving strength $|U_{\bm{n}}|$.

The second relevant parameter is the ratio $\hbar\omega/E_L$ between the
photon energy and the ballistic Thouless energy $E_L=\hbar v_F/L$, i.e.,
the energy scale associated with the length $L$ of the driven region,
with $v_F$ the Fermi energy. This scale governs the formation of
Fabry-Perot scattering resonances (different from the
polarization-dependent resonance to be identified below), so that
$\hbar\omega/E_L$ provides a measure for the number of resonances within
a window $\hbar\omega$. For $\hbar\omega/E_L\gg 1$, evanescent modes are
densely spaced, which means that the quantum pumping mechanism by these
modes is maximally efficient.\cite{SanJose2011a}

We concentrate on the regime of universal and efficient quantum pumping
of $W\gg L$, $p\ll 1$ and $E_L\gg \hbar \omega$. The pumped current \eqref{current} then becomes independent of $L$,
and the response to the driving is proportional to the effective driving
strength $p$, Eq.~\eqref{eq:p2}. Extracting this factor and the scales of
the energy and momentum integrals in the current formula, we obtain a
characteristic current
\begin{equation}
\label{I0laser}
I_0 = \frac{e}{h}\left(\frac{eE_0v_F}{2\omega}\right)^2
\frac{W}{\hbar v_F }.
\end{equation}

A laser in the visible red, with wavelength of $\sim 850$\,nm, 1\,mW
output power and $1\,\mu$m spot size gives a typical coupling energy
$|U_{\bm{n}}|=0.1\,\mathrm{meV}$, and a photon energy
$\hbar\omega=0.7$\,eV. Hence $p\approx 10^{-9}$, and
$\hbar\omega/E_L\approx 2200$, such that the conditions for efficient
pumping are fulfilled. The characteristic current is then $I_0\sim
0.7$\,nA. For a far infrared laser with a wavelength of $50\,\mu$m, a
power of 1\,mW and a spot radius of $50\,\mu$m, we have $p\approx 7\cdot 10^{-6}$
and $\hbar\omega/E_L\approx 1900$, and one may achieve much larger pumped
currents of the order $I_0\sim 50$\,nA.

\begin{figure}
\centering
\includegraphics[width=0.8\columnwidth]{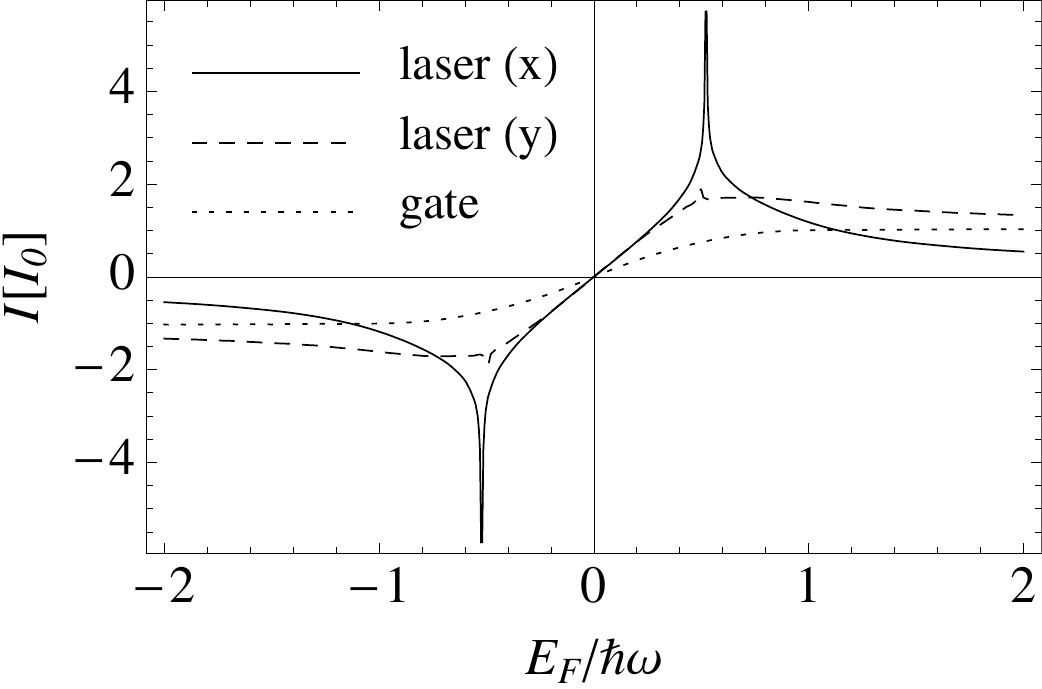}
\caption{Total current driven across a wide and short graphene pump,
as a function of Fermi energy (measured from the Dirac point in the
static sheet) for various driving mechanisms as indicated in the legend.
}
\label{fig:I}
\end{figure}

The precise value of the pumped current can be obtained by wave matching
of modes with fixed transverse momentum $k_y$ and subsequent integration
of the contributions of these modes. The resulting dependence on the
Fermi energy (measured from the Dirac point in the static sheet) is shown
in Fig.~\ref{fig:I}. As a consequence of particle-hole symmetry in the
Dirac Hamiltonian, the pumped current is antisymmetry and vanishes at the
Dirac point. Far from the Dirac point, i.e., for strongly doped graphene
with $|E_F| >\hbar\omega$, the current saturates at a value of the order
$I_0$. Thus, Eq.~\eqref{I0laser} indeed generally reflects the magnitude
of the achievable pumped current. For a laser polarized in transport
direction $x$, however, the pumped current displays a resonant feature at
$|E_F|=\hbar \omega/2$, where $I\gg I_0$. The singularity is logarithmic
and grows as the ratio $\hbar\omega/E_L$ is increased. Circular
polarization (not shown) produces a very similar resonant response. In
contrast, for transverse polarization (along $y$), we witness only a
small hump, while driving by an ac gate voltage does not display any
resonant features.\cite{SanJose2011a}

\begin{figure}
\centering
\begin{tabular}{cc}
\includegraphics[width=0.48 \columnwidth]{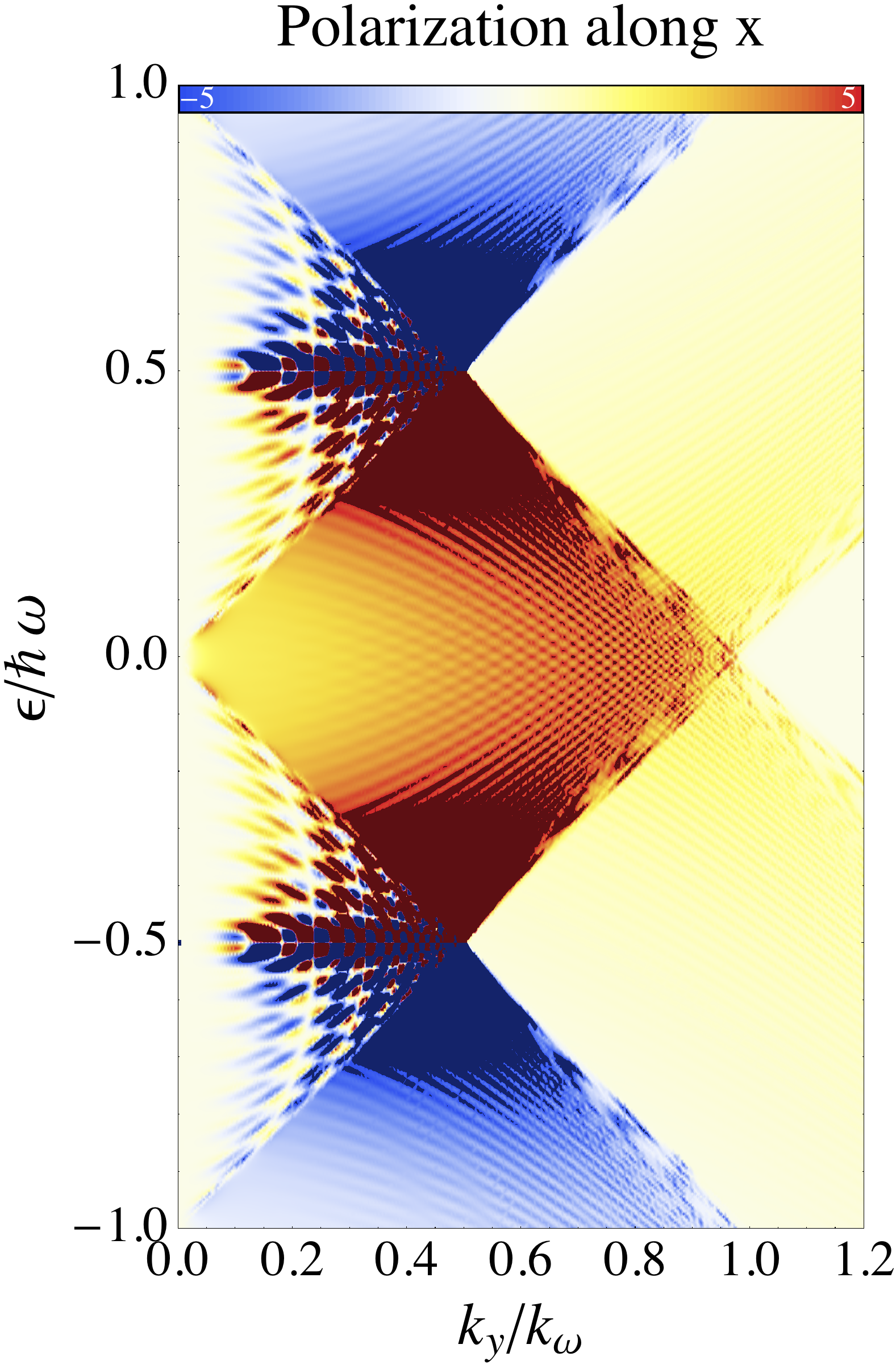} &
\includegraphics[width=0.48 \columnwidth]{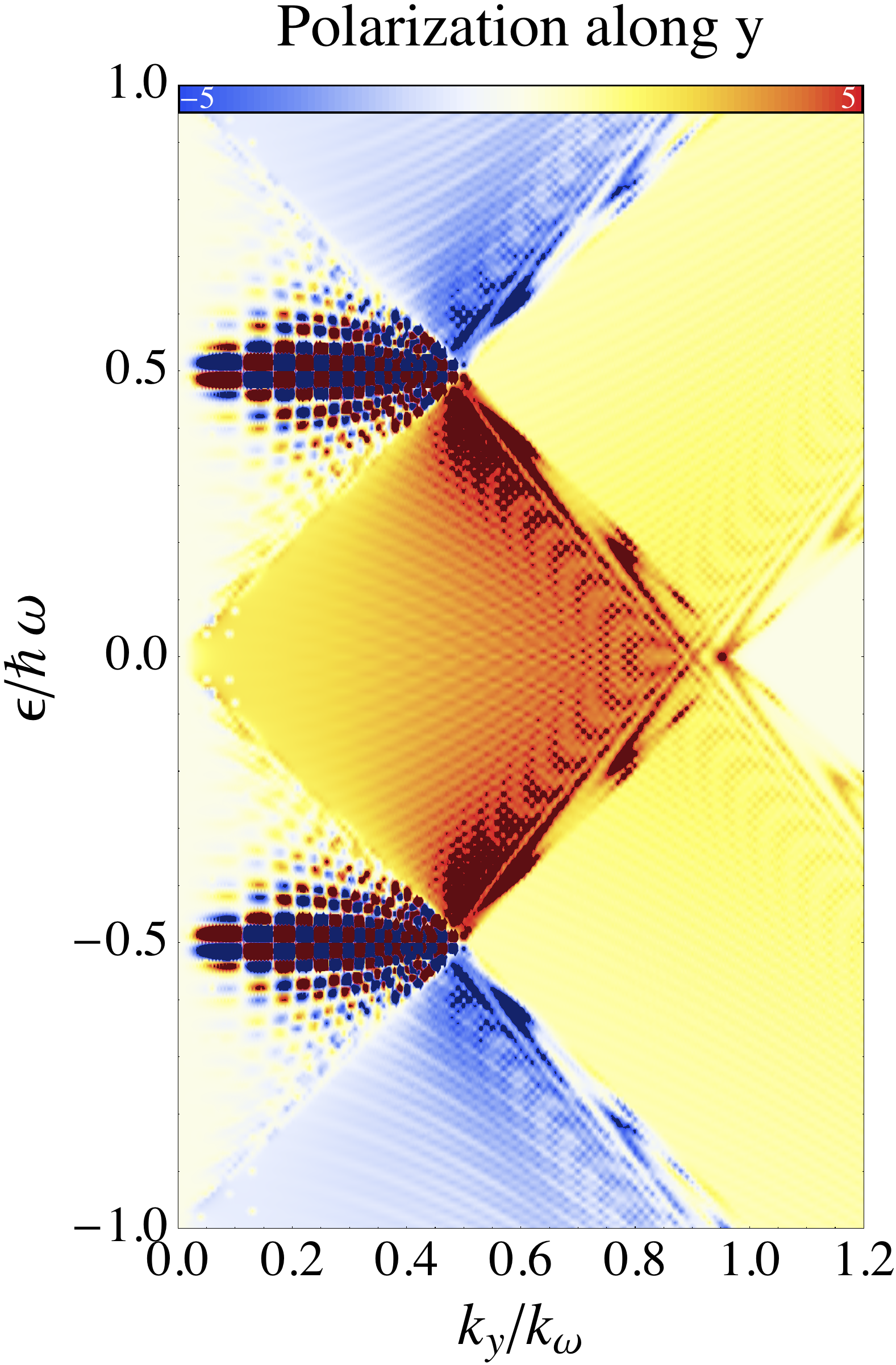} \\
\end{tabular}
\caption{Normalized net transmission $\Delta T/p$, with $p\ll 1$ and
$\hbar\omega/E_L=50$, for laser-driven graphene with polarization in
$x$ (longitudinal) and $y$ (transverse) directions.  The transverse
momentum is measured in units of $k_\omega\equiv \hbar\omega/v_F$. The
color scale (top bar) corresponds to positive current (left to right)
in red, and negative in blue.
}
\label{fig:DeltaT}
\end{figure}

This enhancement of the pumped current for $x$ polarization arises from a
resonant coupling between the pseudo-spin states at $E_F=\pm\hbar
\omega/2$. This is revealed by inspecting the transmission imbalance
$\Delta T(\epsilon,k_y)$, presented in Fig.~\ref{fig:DeltaT}. A
substantial pumping response is obtained in a diamond-shaped area, whose
borders are defined by the transverse momenta at which modes in the
sidebands $n=0,\pm 1$ turn from propagating to evanescent. This
delineates the evanescent-mode pumping mechanism which also dominates the
response to an ac gate voltage.\cite{Prada2009a, SanJose2011a} Notably,
however, for laser driving a strong pumping response also arises from the
sub-harmonic resonances at energies $\epsilon\approx \pm\hbar\omega/2$.
These resonances extend into the range of propagating incoming electrons
with $\hbar v_F k_y<\epsilon$. In the case of $x$-polarization the
resonance evolves into a singular peak around the tips of the evanescent
diamond, at $v_F k_y=\omega/2$.

This additional feature in the pumping response, which is absent for
driving by ac gate voltages, reveals a fundamental feature of the
coupling of laser fields to the chiral charge carriers in graphene. For
fixed wavevector $\bm{k}$, the static Hamiltonian in the pump, $H_0=\hbar
v_F\bm{k}\cdot\bm{\sigma}$, possesses propagating eigenstates
$|\bm{k},s\rangle$ with band index $s=\pm 1$, energies
$\epsilon_{\bm{k},s}=s \hbar v_F|\bm{k}|$ and pseudospin parallel to
$s\bm{k}$. A laser with frequency $\omega$ induces resonant transitions
between any two such states with the same $k_y$, and energy
difference $|\epsilon_{k_x k_y,s}-\epsilon_{k_x' k_y,s'}|=\hbar\omega$.
The transition probability $P= |\langle
k_x',k_y,s'|\bm{\sigma}\cdot\bm{n}|k_x,k_y,s\rangle|^2$ depends on the
matrix element of the normalized driving operator
$U_{\bm{n}}/|U_{\bm{n}}|=\bm{\sigma}\cdot\bm{n}$. This probability becomes
maximal, $P=1$, if $\bm{k},\bm{k'}\perp \bm{n}$, and $s\bm{k}=-s'\bm{k}$.
This resonant condition can be satisfied for $\bm{n}$ in the
$x$-direction, when $v_F\bm{k}=(0,\omega/2)$ and $s=-s'$, which is
manifested as a large pumping response at $|\epsilon|\approx\hbar
v_Fk_y\approx\hbar\omega/2$ (see Fig.~\ref{fig:DeltaT}, left panel). The
response around this point has a constant positive (negative) sign for
$|\epsilon|$ below (above) $\hbar\omega/2$ and thus builds up to a large
peak in the total current. In contrast, for $y$ polarization the response
around $\epsilon=\pm\hbar\omega/2$ oscillates as a function of $k_y$,
which leads to a cancellation of different modes and a small overall
contribution to the total pumped current.

In summary, driving graphene by laser irradiation can lead to a
sizable pumped current, which furthermore is resonantly enhanced when
the polarization is parallel to the transport direction. For lasers in
the visible range, typical currents are $I_0=0.7$\,nA, while the
resonant condition $|E_F|=\hbar\omega/2$ corresponds to a relatively
high doping, which may require chemical functionalization or liquid
electrodes.  Moreover, infrared lasers may be employed to lower the
resonant condition to a much more accessible range of dopings ($|E_F|\sim 12$\,meV), while
at the same time increasing $I_0$ and suppressing possible inelastic
effects due to electron-phonon interactions. The constraint in this
case lies in keeping the laser spot size $L$ below the mean free path
in the flake.

This work was supported by the Spanish Ministry of Economy and
Competitiveness through Grant Nos.\ FIS2008-00124/FIS (P.S.-J.),
FIS2009-08744 (E.P.) and MAT2011-24331 (S.K.).

\bibliography{literature}

\end{document}